\documentclass[reprint,superscriptaddress,amsmath,amssymb,aps,prl,]{revtex4-2}

\usepackage{graphicx}% Include figure files
\usepackage{dcolumn}% Align table columns on decimal point
\usepackage{bm}% bold math
\usepackage{amsmath}

\begin{document}

%\preprint{}

\title{Real-space observation of emergent complexity of \\phase evolution in micrometer-sized IrTe$_2$ crystals}



\author{H. Oike}
\email{oike@ap.t.u-tokyo.ac.jp}
\affiliation{Department of Applied Physics and Quantum-Phase Electronics Centre (QPEC), The University of Tokyo, Tokyo 113-8656, Japan}
\affiliation{RIKEN Center for Emergent Matter Science (CEMS), Wako 351-0198, Japan}

\author{K. Takeda}
\affiliation{Department of Applied Physics and Quantum-Phase Electronics Centre (QPEC), The University of Tokyo, Tokyo 113-8656, Japan}

\author{M. Kamitani}
\affiliation{RIKEN Center for Emergent Matter Science (CEMS), Wako 351-0198, Japan}

\author{Y. Tokura}
\affiliation{Department of Applied Physics and Quantum-Phase Electronics Centre (QPEC), The University of Tokyo, Tokyo 113-8656, Japan}
\affiliation{RIKEN Center for Emergent Matter Science (CEMS), Wako 351-0198, Japan}
\affiliation{Tokyo College, University of Tokyo, Tokyo 113-8656, Japan}

\author{F. Kagawa}
\email{kagawa@ap.t.u-tokyo.ac.jp}
\affiliation{Department of Applied Physics and Quantum-Phase Electronics Centre (QPEC), The University of Tokyo, Tokyo 113-8656, Japan}
\affiliation{RIKEN Center for Emergent Matter Science (CEMS), Wako 351-0198, Japan}


\date{\today}

\begin{abstract}
We report complex behaviors in the phase evolution of transition-metal dichalcogenide IrTe$_2$ thin flakes, captured with real-space observations using scanning Raman microscopy. The phase transition progresses via growth of a small number of domains, which is unlikely in statistical models that assume a macroscopic number of nucleation events. Consequently, the degree of phase evolution in the thin flakes is quite variable for the selected specimen and for a repeated measurement sequence, representing the emergence of complexity in the phase evolution. In the $\sim$20-$\mu$m$^3$-volume specimen, the complex phase evolution results in the emergent coexistence of a superconducting phase that originally requires chemical doping to become thermodynamically stable. These findings indicate that the complexity involved in phase evolution considerably affects the physical properties of a small-sized specimen.
\end{abstract}

%\keywords{}

\maketitle

%\section{}
%\subsection{}
%\subsubsection{}

Condensed matter physics targets many-body systems consisting of atoms/electrons/spins on the order of $\sl{N}$ $\sim$ 10$^3$--10$^2$$^0$. The trajectories of individual constituents in phase space generally show complex and unpredictable behaviors as a result of many-body and/or non-linear interactions. Moreover, it is also unlikely that the exact same microscopic state will be reproduced. Nevertheless, such complexity or chaotic behavior occurring at the microscopic level usually does not cause serious uncertainty about measurements of macroscopic quantities, which are proportional to $\sl{N}$; the underlying complex trajectories are simply observed as fluctuations around the equilibrium state, giving rise to a standard deviation proportional to $\sqrt{N}$. Thus, uncertainty about the measured value decreases in proportion to 1/$\sqrt{N}$, and this scaling ensures the validity of the statistical approach for condensed matter \cite{Landau1980}.

We shall consider phase evolution of a many-body system from metastable to stable states. Because such a phase evolution generally occurs via nucleation and subsequent growth of the final-state domains, the relevant constituents describing the phase evolution are domains. For clarity, let us assume that one nucleation event result in the formation of a final-state domain of volume $\sl{V}$ $\sim$ 1$^3$--10$^3$ $\mu$m$^3$; then, the phase transformation of a 1-mm$^3$ system would include 10$^6$--10$^9$ nucleation events. This number is still macroscopic, and therefore, one can discuss, for instance, the time evolution of the volume fraction of the final state, $\phi$($\sl{t}$), as statistically well-averaged, reproducible behavior \cite{avrami1939kinetics, lifshitz1961kinetics, langer1980kinetics}. In this context, the Johnson-Mehl-Avrami-Kolmogorov (JMAK) equation \cite{avrami1939kinetics}, $\phi$($\sl{t}$$)= 1 - \exp$$(-\sl{kt}$$^n)$ ($\sl{k}$ and $\sl{n}$ are the so-called JMAK parameters), or other smooth functions have been widely used and successfully applied to isothermal phase evolutions in liquids \cite{uhlmann1972kinetic, wuttig2007phase}, alloys \cite{porter2009phase}, and correlated electron systems \cite{Poccia2011oxygenvacancy, oike2015phase, vaskivskyi2015controlling, oike2016interplay, sato2017electronic, sasaki2017crystallization, katsufuji2020nucleation}. Thus, irregular complex phase evolution involved at the microscopic level \cite{jesse2008direct, ricci2015ironvacancy, ricci2020intermittent} does not clearly appear in the macroscopic phase evolution, such as $\phi$($\sl{t}$).

Such a statistical perspective on phase evolution, however, may become less suitable when the system size is small enough that the number of involved nucleation events is not macroscopic. For instance, real-space imaging measurements have revealed that in a colloidal droplet consisting of only $\sim$4,000 particles, a few nucleation events, followed by the growth of the nuclei, are sufficient to complete the crystallization of the entire droplet \cite{gasser2001real}. Such a small number of nucleation events obviously undermines the basis of the statistical perspective, thus implying that complex behavior can appear in a phase evolution of the whole system. The understanding of such complexity involved in the phase evolution is of increasing importance because contemporary sample fabrication techniques have enabled phase control of materials in solid-state nanodevices \cite{wang2012electronics, cao2018unconventional, zhang2019electric}. However, this understanding remains elusive thus far, particularly in electron/spin many-body systems.

%%%%%%%%%%%%%%%%%%%Fig1
\begin{figure*}
\includegraphics{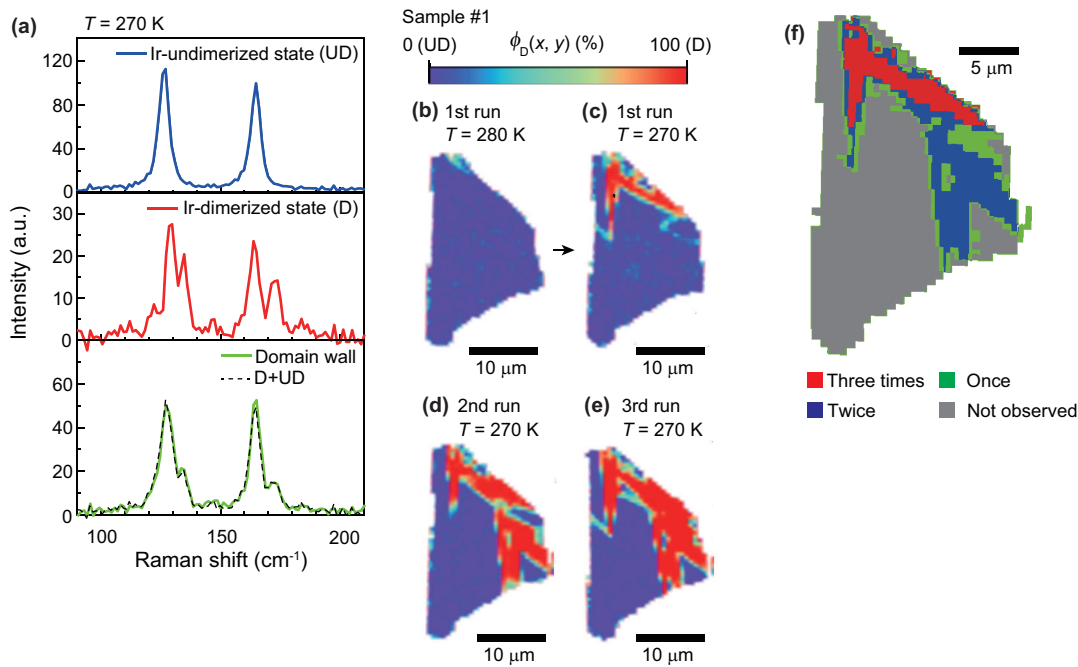}% Here is how to import EPS art
\caption{\label{fig:wide}(a) Raman spectra obtained from an IrTe$_2$ thin plate. The two peaks at $\sim$128 cm$^-$$^1$ and $\sim$165 cm$^-$$^1$ in the Ir-undimerized (UD; top panel) phase split into multiple peaks because of symmetry lowering occurring upon transition to the Ir-dimerized (D; middle panel) phase \cite{lazarevic2014probing, glamazda2014proximity}. The Raman spectrum at (or near) a domain wall (DW; bottom panel) can be reproduced by superposition of the Raman spectra of the UD and D phases (broken line). (b), (c) Domain images at 280 (b) and 270 K (c) upon cooling from 300 K. The spatial distribution of the spectral weight of the D state is represented by the color of each pixel, representing a domain structure consisting of the UD (blue) and D (red) phases. (d), (e) Domain images at 270 K in the second (d) and third (e) cooling runs from 300 K. (f) Superposed image of the three initial-domain images shown in (c)--(e). The color of each pixel represents the number of times―0 (gray), 1 (green), 2 (blue), or 3 (red)―that the Raman spectrum of the D phase was observed in the three repeated sequences.}
\end{figure*}
%%%%%%%%%%%%%%%%%%%Fig1

To examine such variable phase evolution, we targeted the transition-metal dichalcogenide IrTe$_2$. This material exhibits a first-order transition from the Ir-undimerized (UD) state (trigonal, $\sl{P}$$\overline{3}$m1) to the Ir-dimerized (D) state (triclinic, $\sl{P}$$\overline{1}$) at 280--284 K \cite{pascut2014dimerization, eom2014dimerization}. When Pd/Pt/Cu/Rh is doped into IrTe$_2$, the UD state is stabilized down to the lowest temperature, thereby exhibiting a superconducting transition with $\sl{T}$$_\text{c}$ $\sim$ 3.0 K (optimal) \cite{yang2012charge, pyon2012superconductivity, kamitani2013superconductivity, yu2018phase}. By exfoliating an IrTe$_2$ bulk crystal, a small-sized thin-flake sample can easily be prepared, and this material is therefore suitable for studying the complexity involved in phase evolution. Moreover, an IrTe$_2$ thin flake (for instance, a sample volume $\sl{V}$ of $\sim$10--20 $\mu$m$^3$) has been found to exhibit signatures of superconductivity with $\sl{T}$$_\text{c}$ $\sim$ 3.1--3.3 K (ref. \cite{yoshida2018metastable, oike2018kinetic}. See also Fig.~4(a)) even without performing chemical doping; thus, determining how the complexity of the phase evolution affects the lowest-temperature state in a small-sized specimen is also interesting. This issue will also be discussed in this Letter.

In this study, we performed real-space imaging on the entire area of IrTe$_2$ thin plates by using scanning Raman microscopy, with particular interest in the phase evolution. Low-temperature scanning Raman microscopy was performed with a commercially available confocal Raman setup (attoRAMAN, attocube). A low-temperature-compatible objective lens (NA = 0.82, LT-APO 532-RAMAN, attocube) was placed in a cryostat and located 500 $\mu$m above the sample surface, enabling low-temperature Raman imaging with a spatial resolution of $\sim$300 nm (the wavelength of the excitation laser is 532 nm). The Raman signal was collected in the backscattering configuration and detected by a charge-coupled spectrometer with a grating of 1,800 lines mm$^-$$^1$ (for details, see \cite{supple}). The probing depth was estimated to be $\approx$ 10 nm by measuring how the signal intensity for the Si substrate underneath the IrTe$_2$ thin flakes varies with the flake thickness (Fig.~S1 \cite{supple}). 

The typical Raman spectra of the UD and D phases are shown in Fig.~1(a), in the top and middle panels \cite{lazarevic2014probing, glamazda2014proximity}, respectively. The Raman spectrum was measured at each pixel during scanning, and it almost exclusively belonged to either the UD or D phase. Near the domain walls (DWs) separating the two phases, a distinct Raman spectrum was observed, but it was successfully reproduced by superposition of the Raman spectra of the two phases [Fig.~1(a), bottom panel]. Real-space imaging of the two-phase coexistence was thus constructed \cite{supple}. For extracting the complexity of the phase evolution, we focused on whether the results were reproducible for the same sample and whether different propensities were observed for different sample sizes. In the following experiments, we chose the fully UD state at 300 K as the initial state and performed a specific measurement sequence.

We first examined the complexity involved in the nucleation process by scrutinizing the appearance of the initial D domain upon cooling from 300 K. The same measurement sequence was repeated three times for an IrTe$_2$ thin plate (sample \#1: perimeter $\sl{l}$ $\approx$ 77 $\mu$m, thickness $\sl{d}$ $\approx$ 480 nm, and volume $\sl{V}$ $\approx$ 160 $\mu$m$^3$), and the first D domain appeared at a temperature between 270 and 280 K every time [for instance, see Figs.~1(b),(c)]; thus, the transition-onset temperature was roughly reproducible. To gain insight into the reproducibility of the initial domain, we compared the three images obtained at 270 K in each sequence run [Figs.~1(c)--(e)]. In Fig.~1(f), we superposed Figs.~1(c)--(e) and produced a color representation of the frequency of the D domain found at a given pixel. The probability distribution of finding the initial D domain is weighted to certain areas of the specimen, indicating that nucleation sites activated at 270 K are heterogeneously imprinted on this specimen. We note that the fully UD state at 280 K appears to be homogeneous with respect to the Raman shift (Fig.~S2 \cite{supple}), and thus detection of the heterogeneous nucleation sites is an issue beyond the experimental resolution. 

Such a heterogeneous nucleation mechanism is expected to emphasize the variations in nucleation probabilities among different samples, particularly when the sample size is small. For instance, Fig.~1(f) implies that if sample \#1 was further broken into smaller pieces, then whether the D domain appears at 270 K would depend on the piece. As inferred in this thought experiment, whether a selected piece contains a nucleation site that has sufficient nucleation probability at a considered temperature is thus increasingly unpredictable as the sample size decreases. This complexity involved in the nucleation process indicates that a deeply supercooled UD state with no D domain may be found in a small sample. By contrast, such a situation is not expected to be probable in a large sample, which is thought to invariably contain activated nucleation sites somewhere in the sample. 

To verify this issue, we performed consecutive domain imaging for a thermal cycle of 300 $\rightarrow$ 200 $\rightarrow$ 300 K for sample \#1 and a smaller sample \#2 ($\sl{l}$ $\approx$ 49 $\mu$m, $\sl{d}$ $\approx$ 160 nm, and $\sl{V}$ $\approx$ 22 $\mu$m$^3$), and the results are displayed in Fig.~2(a) and Fig.~S3 \cite{supple}, respectively. The temperature evolution of the D domain volume fraction [Fig.~2(b)] shows that the temperature at which the first D domain appears is appreciably lower in sample \#2 than in sample \#1, consistent with the above expectation and the transport measurements in the literature \cite{oike2018size}. We also note that the D-to-UD transition temperatures observed upon heating are nearly the same as that of the bulk crystal \cite{eom2014dimerization, oike2018size}, implying that the thermal equilibrium phase diagrams of the two samples are affected only weakly by sample smallness at the level considered. Thus, the observed deep-supercooling phenomena in the smaller sample (\#2) can be accounted for primarily by the absence of active nucleation sites, rather than by a decrease in the equilibrium transition temperature. For an even smaller sample, the D-to-UD transition temperature is appreciably lower than that of the bulk sample (Fig.~S4 \cite{supple}; see also \cite{park2020superconductivity}), and we excluded such a small sample in the present study (for more details, see \cite{supple}).

%%%%%%%%%%%%%%%%%%%Fig2
\begin{figure}
\includegraphics{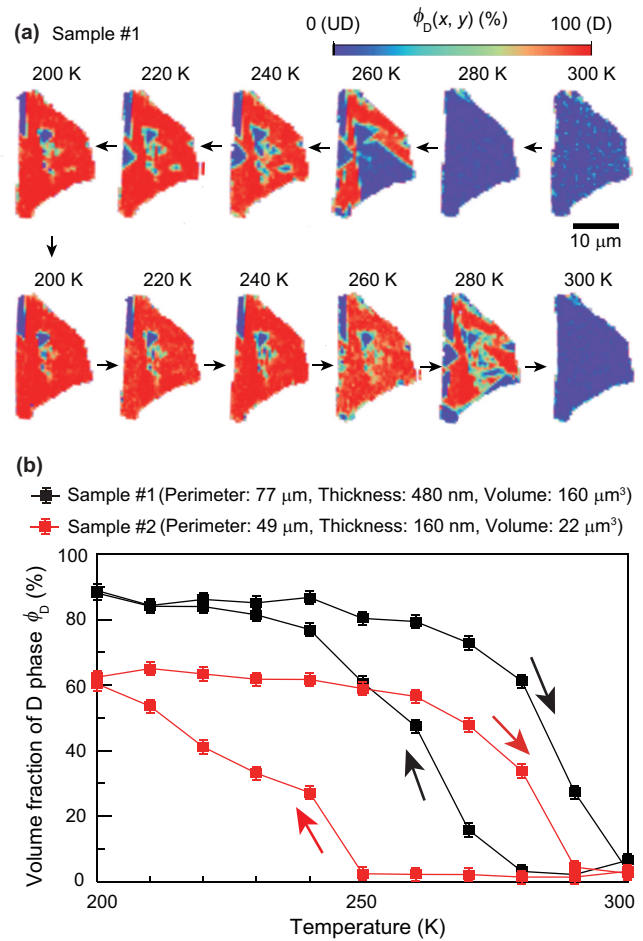}
\caption{\label{} (a) Consecutive domain images of sample \#1 for a thermal cycle of 300 $\rightarrow$ 200 $\rightarrow$ 300 K. For those of sample \#2, see Fig.~S3 \cite{supple}. (b) Temperature evolution of the volume fraction of the charge-ordered phase in samples \#1 and \#2.}
\end{figure}
%%%%%%%%%%%%%%%%%%%Fig2

To understand the size-dependent supercooling phenomena in more detail, it may be important to consider nucleation at the perimeter. The close inspection of Fig.~2(a) and Fig.~S3 \cite{supple} reveals that during the structural transition, both on heating (D to UD) and cooling (UD to D), new structural domains tend to appear near the perimeter of the flakes. This preferred nucleation at the perimeter may be explained by considering that nucleation near the free boundaries is helpful in reducing the interfacial strain energy associated with the DW area \cite{porter2009phase} and/or the bulk strain energy associated with the volume difference between the high- and low-temperature phases \cite{bai2005bulkelastic}. Thus, the complexity regarding the heterogeneous nucleation process in IrTe$_2$ thin flakes is likely to be more sensitive to the microscopic details of the perimeter, rather than those of the bulk.

We next examined the complexity involved in the growth process of the D domain. Isothermal consecutive domain imaging was performed for sample \#1 at 270 K, and the sequence was repeated two times [Figs.~3(a),(b)]. The time evolutions of the volume fractions of the D phase, $\phi$$_\text{D}$($\sl{t}$), were thus derived as shown in Fig.~3(c). Overall, in the first run, the volume fraction gradually increased with time, but it exhibited an abrupt increase when $\sim$24 hours passed after beginning the observations. Interestingly, in the second run, an abrupt increase with a larger magnitude was observed at an earlier elapsed time, even though we nominally used the same measurement procedure from 300 K. Obviously, the $\phi$$_\text{D}$($\sl{t}$) profile obtained in each sequence is not reproducible, and moreover, it cannot be described by a function as smooth as the statistical model predicts. Thus, the complexity involved in the growth process is manifested in the $\phi$($\sl{t}$) profile, a situation that is not expected in a sufficiently large sample. The close inspection of two successive domain images [at 24.0 h and 24.6 h in Fig.~3(a), or at 3.6 h and 4.3 h in Fig.~3(b)] suggests that the intermittent pinning and depinning of DWs underlie the observed complexity. 

%%%%%%%%%%%%%%%%%%%Fig3
\begin{figure}
\includegraphics{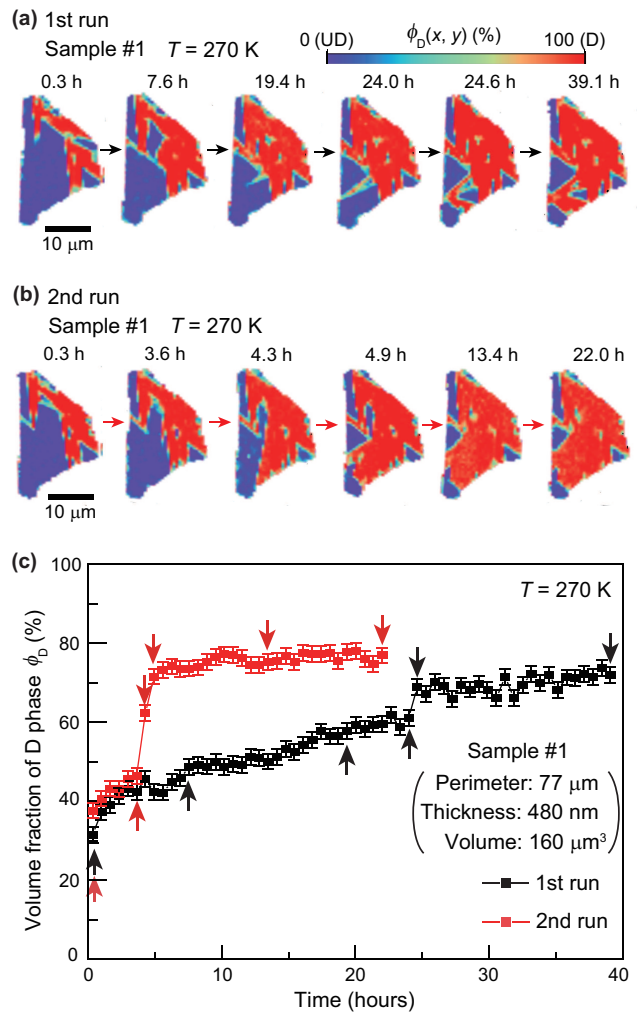}
\caption{\label{} (a), (b) Isothermal time evolution of the domain images at 270 K after cooling from 300 K for twice repeated sequences. (c) Isothermal time evolution of the volume fraction of the charge-ordered phase at 270 K after cooling from 300 K. The times corresponding to the domain images shown in (a), (b) are indicated by arrows in (c).}
\end{figure}
%%%%%%%%%%%%%%%%%%%Fig3

Finally, we discuss how the complexity involved in the phase evolution is linked to the superconductivity emerging in the nondoped small-sized IrTe$_2$. As shown in Fig.~4(a), the nondoped IrTe$_2$ specimens of $\sl{V}$ $\sim$ 10--20 $\mu$m$^3$ are prone to exhibit signatures of superconductivity, whereas those of $\sl{V}$ $>$ 100 $\mu$m$^3$ are not \cite{oike2018kinetic}. We performed domain imaging at 2 K and found that, reflecting the size-dependent transport properties, samples \#1 and \#2 were in different domain states. Sample \#1 exhibited a fully D state [Fig.~4(b)], as expected from the phase diagram of the bulk sample; thus, the complex phase evolution does not appear to have a significant impact on the lowest-temperature state in sample \#1 after all. By contrast, the smaller sample (\#2) exhibited pronounced coexistence of the D and UD phases [Fig.~4(c)], implying that the complex intermittent depinning of the DWs ceased during cooling to 2 K in sample \#2. The resulting quenched UD phase is thus thought to be the origin of superconductivity in the nondoped IrTe$_2$ thin flakes of $V$ $\sim$ 10--20 $\mu$m$^3$. In fact, in doped bulk IrTe$_2$, superconductivity appears when the UD phase is thermodynamically stabilized down to the lowest temperature \cite{yu2018phase}, and thus it is plausible that the quenched UD phase exhibits superconductivity.

%%%%%%%%%%%%%%%%%%%Fig4
\begin{figure*}
\includegraphics{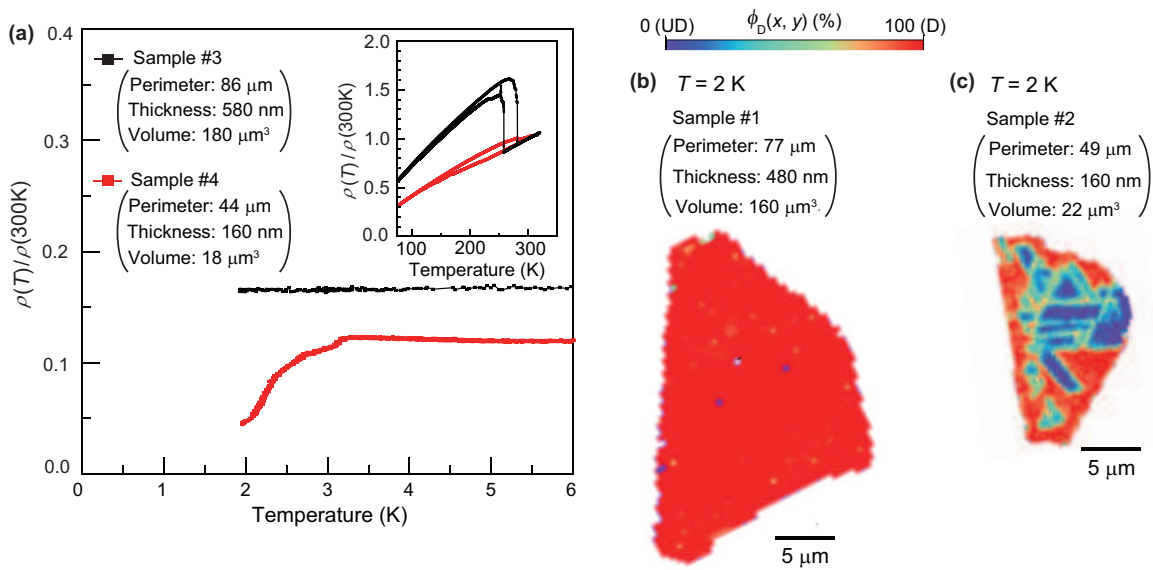}% Here is how to import EPS art
\caption{\label{fig:wide} (a) Temperature-resistivity profiles in samples \#3 and \#4 near the lowest temperature. Inset: Temperature-resistivity profiles near the first-order phase transition. The sample perimeter, thickness, and volume are 86 $\mu$m, 580 nm, and 180 $\mu$m$^3$ for sample \#3 and 44 $\mu$m, 160 nm, and 18 $\mu$m$^3$ for sample \#4. (b), (c) Domain images at 2 K of samples \#1 (b) and \#2 (c). Note that although the sample geometries are close to each other [samples \#1 (\#2) and \#3 (\#4)], the sample used for the resistivity measurement (a) is different from those used for scanning Raman microscopy [(b) and (c)]. Cooling from 300 to 2 K was performed several times, and the domain images for repeated cooling are shown in Fig.~S5 \cite{supple}. When the sample-holder temperature was cooled to 2 K, the excitation laser was switched off, and hence, the sample temperature was equilibrated with the sample-holder temperature during cooling. At the lowest temperature, the specific heat of the sample became small, and thus the local temperature at the focused laser spot may be increased by on the order of tens of degrees K. Nevertheless, we confirmed that the applied laser intensity did not induce a phase change between the UD and D states (Fig.~S6 \cite{supple}), and therefore, it is safe to say that the domain pattern is unaffected by the laser power used for scanning (5 mW). }
\end{figure*}
%%%%%%%%%%%%%%%%%%%Fig4

The scenario in which the complexity of the phase evolution underlies the emergent lowest-temperature state in sample \#2 is further supported by the fact that the phase-coexistence pattern varies with repeated cooling from 300 K (see Fig.~S5 \cite{supple}). In sample \#1, by contrast, the phase transformation can invariably be completed during cooling to 2 K (Fig.~S5 \cite{supple}), again indicating that the complexity of the phase evolution does not significantly affect its lowest-temperature state. This contrasting behavior implies that the DWs in the smaller sample are more prone to be pinned compared with those of the larger sample. In fact, an IrTe$_2$ thin flake often shows a broadened and incomplete first-order phase transition (for instance, see the Fig.~4(a) inset and the literature \cite{yoshida2018metastable, oike2018kinetic}), also suggesting low DW mobility. The specific underlying mechanism in IrTe$_2$ is not clear from the present experiment, but we note that a lower DW mobility is expected for a thinner sample from a simple balance-of-force equation that takes into account pinning due to both bulk and surface inhomogeneities \cite{lebedev1994surface, tagantsev2010domains}. This $\lq$$\lq$surface pinning size effect" is likely to play a role in rendering a small sample susceptible to the complexities of phase evolution. 

We have shown that in contrast to the case for bulk samples, complex behaviors dominate the phase evolution of micrometer-sized thin-flake samples, resulting in pronounced variations for the selected specimen and the repeated measurement sequence. These results highlight the inadequacy of the statistical approach when considering a small system exhibiting a first-order phase transition. Additionally, the complex behaviors in phase evolution suggest that phenomena deviating from the properties in the corresponding bulk sample may appear, such as the emergence of superconductivity that would otherwise require chemical doping. Note that, in an even smaller sample, the dominance of the bulk free energy would eventually collapse due to increased surface contributions, potentially resulting in a new electronic state. To achieve a comprehensive understanding of phase transformations in micro- and nanoelements, consideration of both the complexity of phase evolution and surface contributions to the total free energy would thus be important.

% Surround figure environment with turnpage environment for landscape
% figure
% \begin{figure}
% \end{figure}

% tables should appear as floats within the text
% The ruledtabular enviroment adds doubled rules to table and sets a
% reasonable default table settings.
% Use the table* environment to get a full-width table in two-column
% Add \usepackage{longtable} and the longtable (or longtable*}
% environment for nicely formatted long tables. Or use the the [H]
% placement option to break a long table (with less control than 
% \begin{table}%[H] add [H] placement to break table across pages
% \begin{ruledtabular}
% Lines of table here ending with \\
% \end{tabular}
% \end{ruledtabular}
% \end{table}

% Surround table environment with turnpage environment for landscape
% table
% \begin{table}
% \begin{ruledtabular}
% \end{tabular}
% \end{ruledtabular}
% \end{table}

%\appendix
%\section{}

\begin{acknowledgments}
H.O. and F.K. thank Y. Ogimoto, H. Suwa, and K. Shibata for their valuable discussions. This work was partially supported by JSPS KAKENHI (Grant Nos. 18K13512, 20K14410, 18H01168, and 21H04442) and JST CREST (Grant No. JPMJCR1874).
\end{acknowledgments}

%\bibliography{Oike}
%apsrev4-1.bst 2019-01-14 (MD) hand-edited version of apsrev4-1.bst
\end{document}